\documentclass[a4paper]{article}
\usepackage[pdftex]{graphicx}
\usepackage{amsmath}
\usepackage{amssymb}
\usepackage{amsthm}
\usepackage[utf8]{inputenc}
\usepackage[T1]{fontenc}
\usepackage{hyperref}
\usepackage{smartref}
\usepackage{todonotes}
\usepackage{fullpage}
\usepackage{epstopdf}
\usepackage{subcaption}

\usepackage[toc,page]{appendix}

\usepackage{color}

\renewcommand\vec{\boldsymbol}

\newcommand{\scal}[2]{\langle #1,#2\rangle}

\newcount\colveccount
\newcommand*\colvec[1]{
        \global\colveccount#1
        \begin{pmatrix}
        \colvecnext
}
\def\colvecnext#1{
        #1
        \global\advance\colveccount-1
        \ifnum\colveccount>0
                \\
                \expandafter\colvecnext
        \else
                \end{pmatrix}
        \fi
}

\begin{document}

\title{Analytical graphic statics}
\author{Tamás Baranyai}
\maketitle

\abstract
Graphic statics is undergoing a renaissance, with computerized visual representation becoming both easier and more spectacular as time passes. While methods of the past are revived and tweaked, little emphasis has been placed on studying the details of these methods. Due to the considerable advances of our mathematical understanding since the birth of graphic statics, we can learn many interesting and beautiful things by examining these old methods from a more modern viewpoint. As such, this work shows the mathematical fabric joining different aspects of graphic statics, like dualities, reciprocal diagrams and discontinuous stress functions.

\section{Introduction}
Graphic statics was born with the works of Culmann \cite{culmann1875graphische} and Bow \cite{bow2014economics}, and it contained two diagrams, namely the force diagram and the geometrical or form diagram. The force diagram represented magnitudes of forces with lengths of the corresponding line segments, while each line in the form diagram gave the line of action of the corresponding force. In the early days of graphic statics, these two diagrams were connected by algorithmic methods. As structures became more and more complex, this algorithmic approach became too cumbersome, and graphic methods gave way to algebraic ones. With the emergence of computers and their huge visualising power, these algorithmic methods became feasible once again and are proposed as optimizing tools \cite{VANMELE2014104,ALIC201726,hablicsek2019algebraic}.

The next step in the evolution of graphic statics came with Maxwell \cite{maxwell1870}, who gave a 3 dimensional graphical construction to solve a two dimensional truss. His method introduced two new aspects to graphic statics: One was the use of projective geometrical dualities (a symmetric polarity in particular) to establish connection between the form and force diagrams, the other was the notion of a discontinuous stress function corresponding to the discontinuous structure the truss is. These two notions went on to live somewhat distinct evolutionary paths as more research followed. 

The set of usable projective geometrical dualites was quickly expanded with null-polarites (symplectic polarites) by Cremona \cite{cremona1890graphical}. Polarites were considered to be special in this regard by researchers \cite{KONSTANTATOU2018272} until very recently, when it has been shown \cite{2019arXivBaranyai} that any projective duality is usable for this purpose and in infinitely many ways. 

The idea of using a stress function for analysing discontinuous structures was quickly supported by Klein and Wieghardt \cite{klein1904onMaxwell} with some rigorous mathematics, but the next paper came much later. It was in the 1980-s when the topic emerged again, interestingly for spatial trusses first \cite{hegedus84}, then for planar ones \cite{hegedus85}. With the renewed interest a number of works \cite{Mitchell2016partI,McRobie2016partII} described the use of this tool, essentially coining it Airy stress function (similarly to the continuum case). Apart from trusses, a recent pair of papers \cite{williams16,mrobiewilliams19discontinous} tackled the case of planar and spatial frames respectively. 
Yet, even these fairly recent works, while noting that after defining polyhedral stress functions one ought to have moments as function values and mentioning the possible use of projective dualities, did not explain the connection between the proposed elements.\\

What appears to be missing is the explanation of the mathematical fabric tying these concepts together. This paper attempts to do that, with the simplest mathematical tool possible. The tool of choice is the notion of dual spaces associated to vector-spaces, motivated by the modern mathematical approach to projective geometry \cite{faure2013modern}.

\clearpage

\section{The dual nature of forces}
The main point of this paper is that forces can be considered both as vectors and as linear functionals, and graphic statics used both approaches in its diagrams. In order to see this, we will occasionally need projective homogeneous coordinates. Although they are becoming fairly standard, the reader is provided with a brief introduction to them in the Appendix, if necessary.\\

Let us consider a planar body acted on by a planar force system comprised of forces $f_i$ ($i\in \{1,2 \dots n \}$) and let us have an orthogonal $x,y$ coordinate-frame in the plane. 
We can describe any force $f_i$ acting in the plane with the triplet $f_i=(F_{x,i},F_{y,i},M_i)$, where $F_{x,i}$ and $F_{y,i}$ are the projections of the force to the coordinate axes and $M_i$ is the moment of the force with respect to an axis perpendicular to the plane passing through the origin (with respect to the origin for short). 
Let $v_0\in \mathbb{R}^3$ be an arbitrary vector, and let $v_i$ be defined as
\begin{align}
v_i:=v_{i-1}+f_i.
\end{align}
The equilibrium of forces $\sum f_i=0$ ($i\in \{1,2 \dots n \}$) holds precisely if $v_n=v_0$, and this can be graphically represented with a closed, arrow-continuous chain of vectors $f_i$. Although this diagram has not been used often in this 3 dimensional form, the projection of this to the $M=0$ plane is nothing else then the well known "force diagram".\\

Now let $s_i$ be defined as
\begin{align}
s_i:=(-F_{y,i},F_{x,i},M_i)
\end{align}  
and let us note, that the the scalar product $\scal{(q_x,q_y,1)}{s_i}=-q_xF_{y,i}+q_yF_{x,i}+M_i$ is the moment of the force with respect to the point $(q_x,q_y)$. We have embedded our 2D problem into 3D with $z=1$ and we can think of $s_i$ as a linear functional, which can be evaluated at points of the vector-space the planar body is in. Naturally $s_i$ uniquely corresponds to (represents) force $f_i$, and the line of action of force $f_i$ is the intersection of the zero-space of $s_i$ with the plane $z=1$. Let us also note, how the triplets $(q_x,q_y,1)$ can be thought of as carefully chosen representants from the equivalence class $(q_x,q_y,1)_{\sim}$, which is the projective point coordinate form of a point in the $x,y$ plane. In this context the line of action of $f_i$ is precisely the equivalence class $s_{i\sim}$. Note, how given two line representants $s_i$ and $s_j$, the sum $s_i+s_j$ represents a line passing through the intersection point of $s_i$ and $s_j$. The corresponding mechanical property is that the sum of two forces has to pass through the intersection of the two original forces. After this has been pointed out, it is easy to see how this functional formulation lies under the "form diagram" of graphic statics.\\   

Since this paper is about graphic statics, we will want to draw this functional. One way to do it, is to measure up the value $\scal{(q_x,q_y,1)}{s_i}$ over each point $(q_x,q_y,0)$ resulting in plane $-F_{y,i}x+F_{x,i}y,-z,+M_i=0$, or with homogeneous coordinates
\begin{align}
(-F_{y,i},F_{x,i},-1,M_i)_{\sim}\label{eq:rajz}
\end{align}
(alternatives to this will be addressed later, in Section \ref{sec:cr}).

\subsection{The projective connection}
Since a linear combination of functionals is also a functional and a linear combination of plane-representants also represents a plane, we are able to apply a projective duality to the 3 dimensional force diagram introduced earlier and extract the lines of action of the forces involved, as well as create the moment diagram of the structure. Any projective duality is applicable for this purpose, with the appropriate representation of moments. This representation will be addressed later, first a few examples are provided that will serve as a basis the general case can be reduced to.\\

Let us number the forces according to how they follow each other on the body, and create the 3 dimensional force diagram as defined above. The vertices of this diagram are then mapped to planes by a duality, which is describable with an invertible matrix equivalence class $D_{\sim}$ as:
\begin{align}
p_{i\sim}:=(v_i, 1)_{\sim}D_{\sim}. 
\end{align}
Due to the way the force diagram is defined, $p_{i\sim}-p_{i-1\sim}$ represents force $f_i$ and $p_{i\sim}-p_{0\sim}$ represents the resultant of the force system $\sum_i f_i$. Here $i$ may also run on a subset of $i\in \{1,2 \dots n \}$,  depending on the forces we wish to consider the resultant of. The problem that we do not see these planes directly on the dual figure can be solved by observing the relation of the parts forming these linear combinations. The orthogonal projection to the $z=0$ image plane of the intersection of $p_{i\sim}$ and $p_{i-1\sim}$ will be the line of action of $f_i$, while the projection of the intersection of $p_{i\sim}$ and $p_{0\sim}$ will be the line of action of $\sum_i f_i$. Furthermore, the moment diagram of the structure appears between plane $p_{0\sim}$ and $p_{i\sim}$, in the planes the vertical projecting rays running through the structure graze; as the moment the structure is subjected to at any point is caused by the resultant $\sum_i f_i$ at that point.

\subsection{Examples}
\begin{figure}[h]
\begin{subfigure}[t]{0.5\textwidth}
\centering
\includegraphics[scale=0.5]{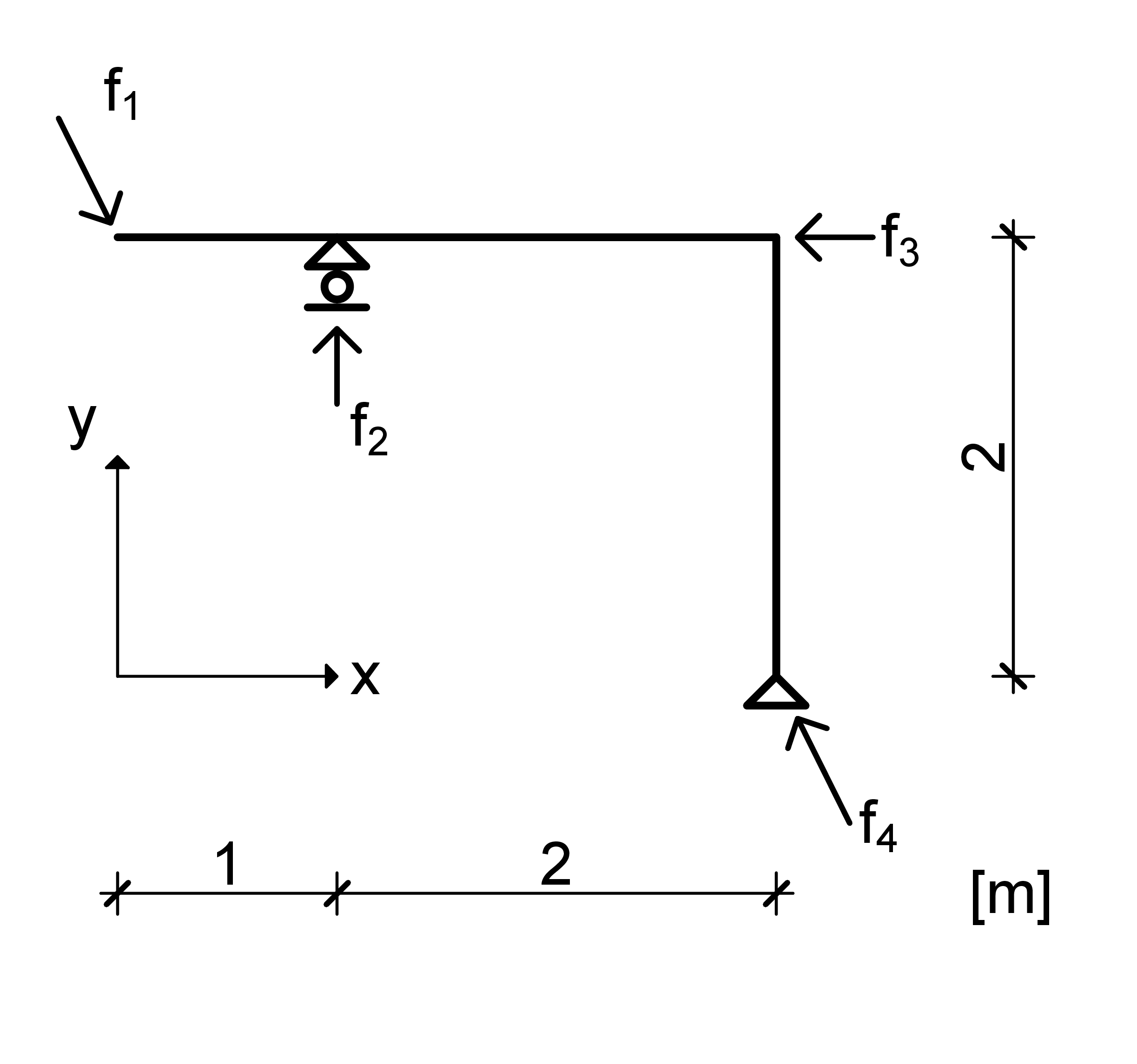}
\caption{Geometry of the first example}\label{fig:rajz1}
\end{subfigure}
\begin{subfigure}[t]{0.5\textwidth}
\centering
\includegraphics[scale=0.5]{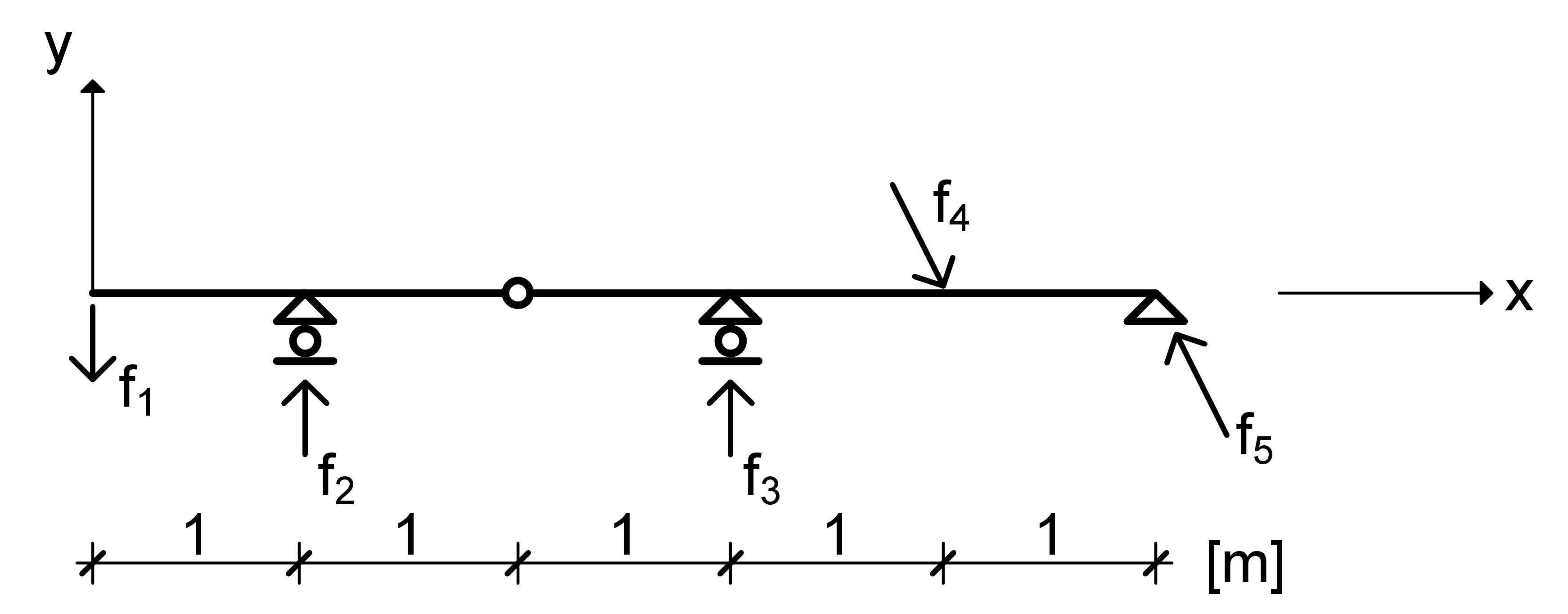}
\caption{Geometry of the second example}\label{fig:rajz2}
\end{subfigure}
\caption{}
\end{figure}

\begin{figure}[h]
\begin{subfigure}[t]{0.5\textwidth}
\includegraphics[width=1\textwidth]{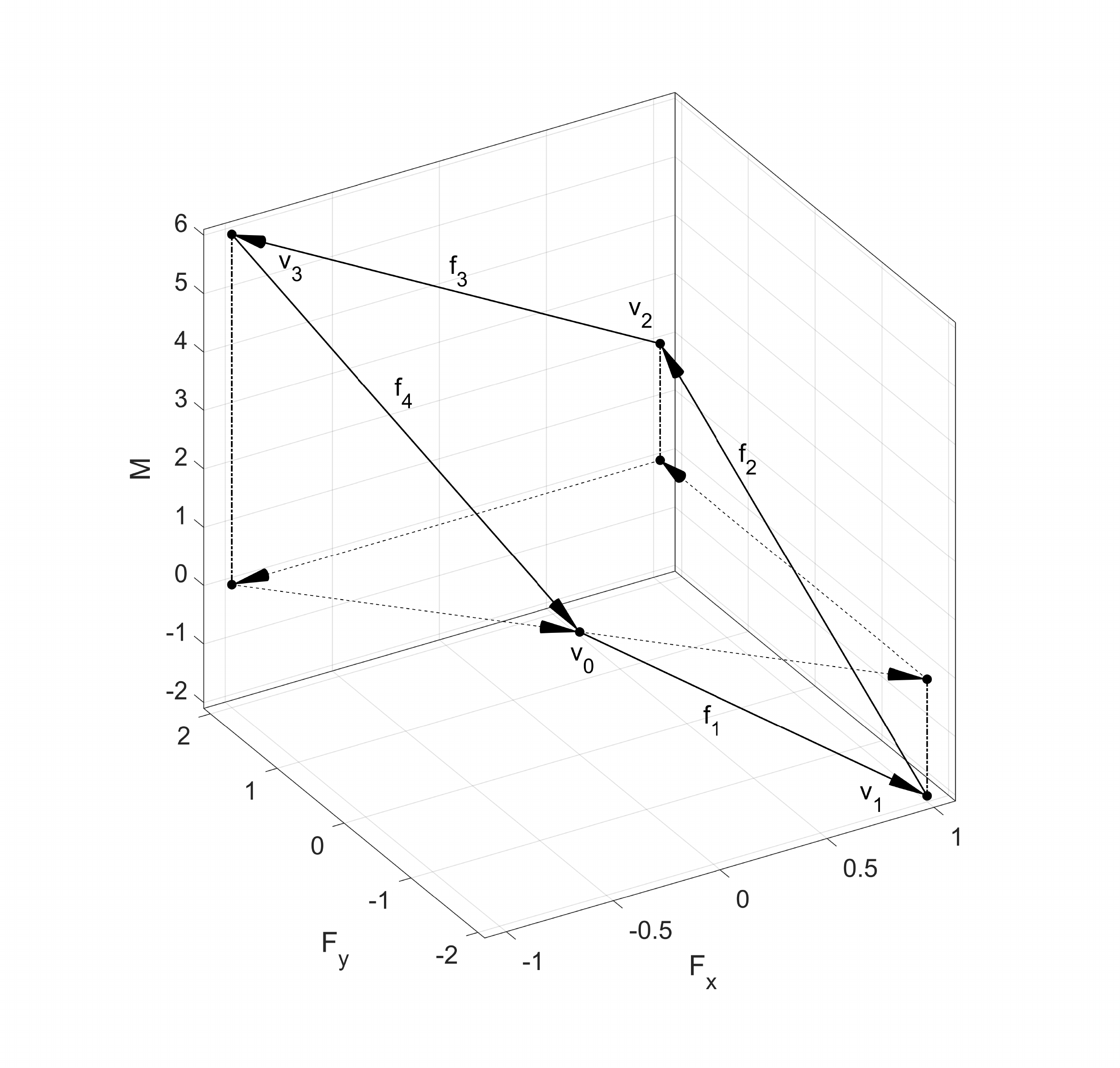}
\caption{Force diagram of the structure shown in Figure \ref{fig:rajz1}.}\label{fig:pl1a}
\end{subfigure}
\begin{subfigure}[t]{0.5\textwidth}
\includegraphics[width=1\textwidth]{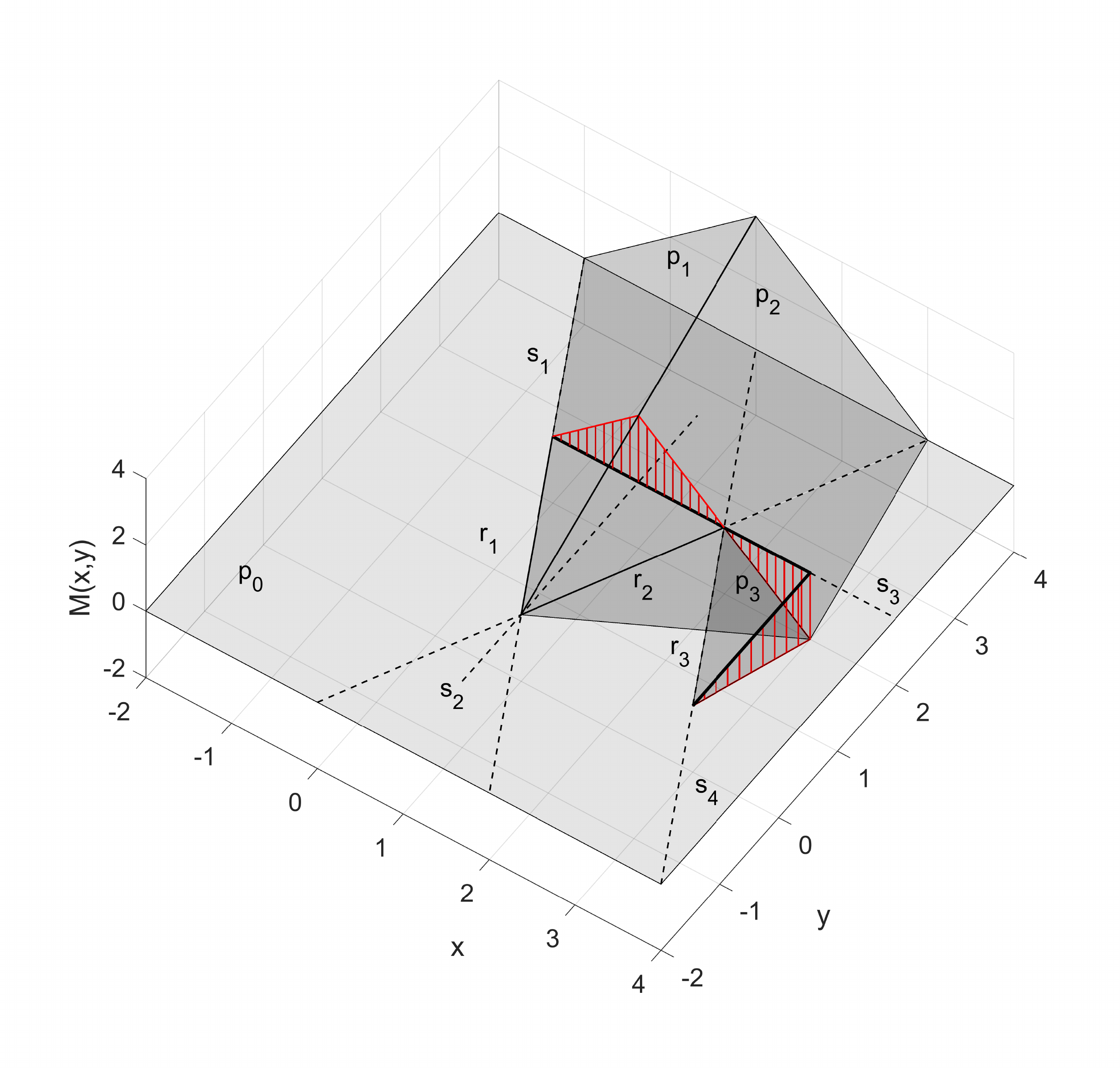}
\caption{Moment functionals evaluated above the structure given in Figure \ref{fig:rajz1}. The moment diagram is shown in red. the lines of actions of the forces are denoted with $s_i$ while the lines of action of the resultant with $r_i$.}\label{fig:pl1b}
\end{subfigure}
\caption{}
\end{figure}

\paragraph{A basic example} is presented in Figure \ref{fig:rajz1}. It is in equilibrium with forces
\begin{align}
f_1=&(1,-2,-2)\\
f_2=&(0,4,4)\\
f_3=&(-2,0,4)\\
f_4=&(1,-2,-6).
\end{align}
Let us pick $v_0=(0,0,0)$ and create the force diagram. It is shown in Figure \ref{fig:pl1a}. We can now use a null-polarity to map $v_{i\sim}$ to $p_{i\sim}$ as
\begin{align}
(F_x,F_y,M,1)_{\sim} \mapsto (-F_y,F_x,-1,M)_{\sim}\label{eq:duC}
\end{align}
which is the drawing of the functionals introduced in \eqref{eq:rajz}. Let us note how $p_{0\sim}$ is the $x,y$ plane, meaning the intersection of $p_{i\sim}$ with it is already the line of action of the resultant without any projection. Furthermore, if the resultant of the forces at a given point of the structure is $\sum_i f_i$, the bending moment the structure is subjected to at that point can be measured vertically  over the point, between $p_{i\sim}$ and the $x,y$ plane. Correspondingly, the projection required to get the lines of action of the forces is an orthogonal projection along the vertical direction (see Figure \ref{fig:pl1b}).    

\begin{figure}[h]
\begin{subfigure}[t]{0.5\textwidth}
\includegraphics[width=1\textwidth]{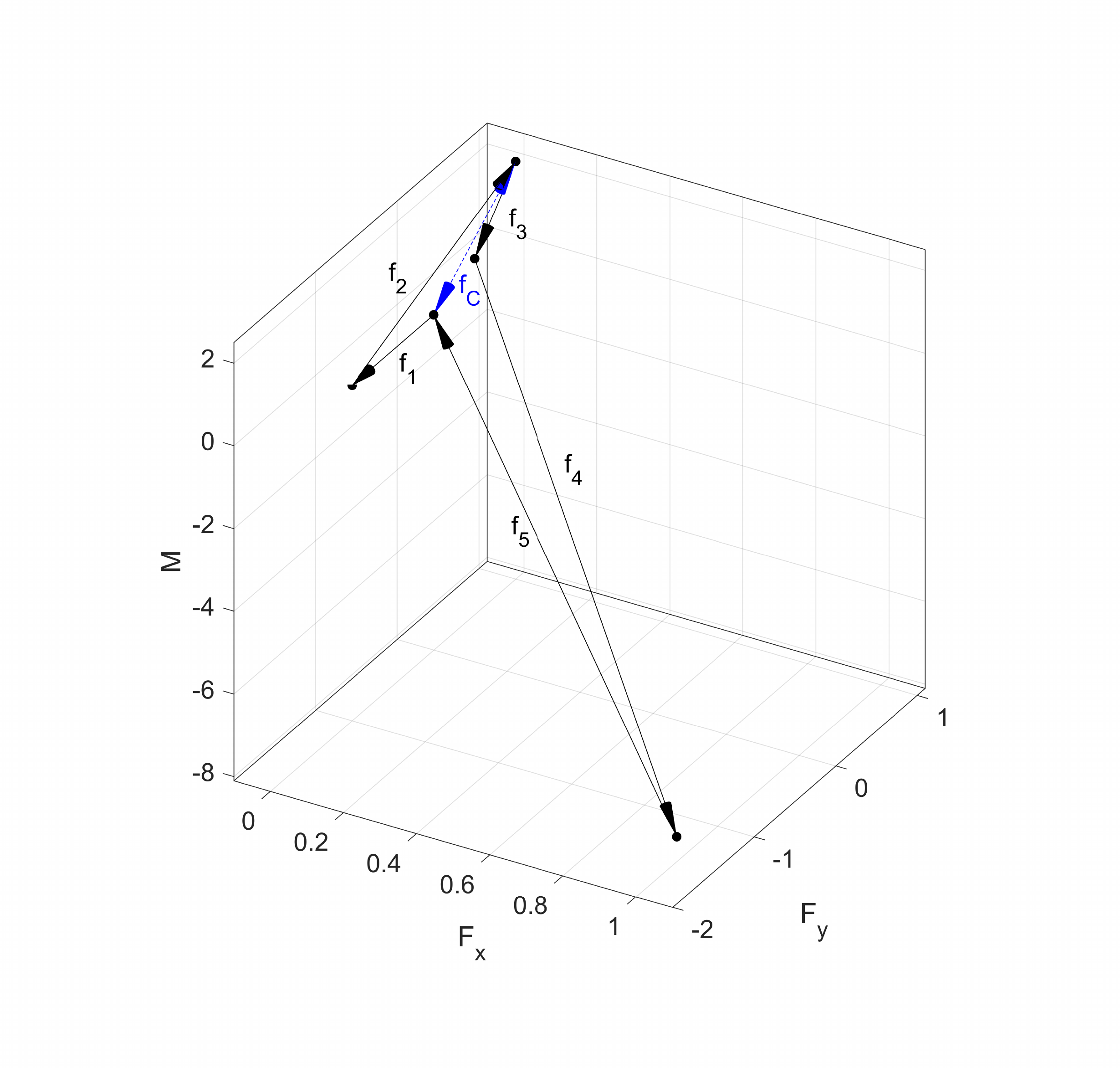}\\
\caption{Force diagram of the structure shown in Figure \ref{fig:rajz2}}\label{fig:pl2a}
\end{subfigure}
\begin{subfigure}[t]{0.5\textwidth}
\includegraphics[width=1\textwidth]{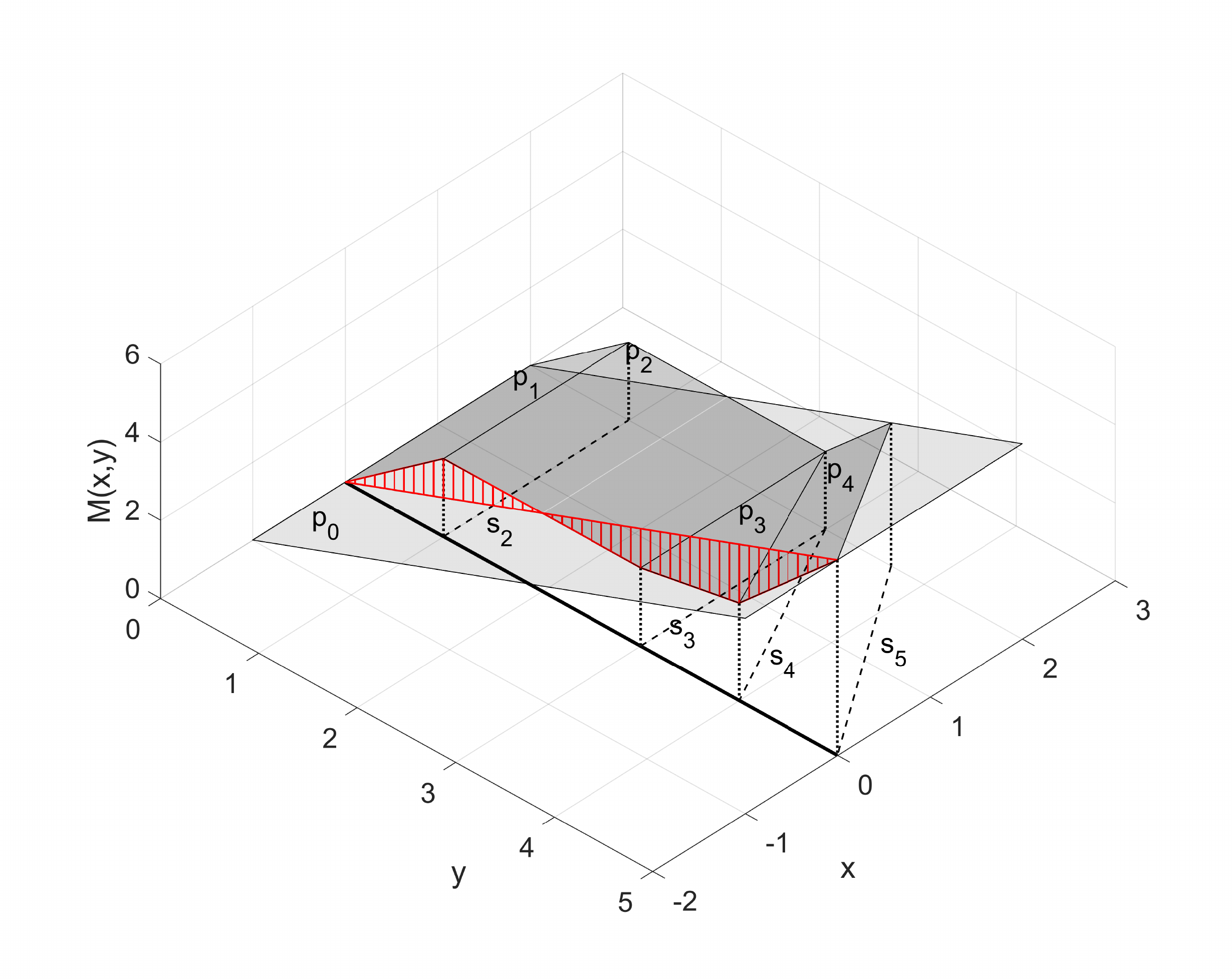}
\caption{Moment functionals evaluated above the structure given in Figure \ref{fig:rajz2}. The moment diagram is shown in red. the lines of actions of the forces are denoted with $s_i$.}\label{fig:pl2b}
\end{subfigure}
\caption{}
\end{figure}

\paragraph{The reference plane for the moment diagram} is not necessarily the $z=0$ plane. Illustrating this, consider the second example presented in Figure \ref{fig:rajz2}. The statical equilibrium is satisfied with forces
\begin{align}
f_1=&(0,-1,0)\\
f_2=&(0,2,2)\\
f_3=&(0,-1/2,-3/2)\\
f_4=&(1,-2,-8)\\
f_5=&(-1,3/2,15/2).
\end{align}
The corresponding force diagram is shown in Figure \ref{fig:pl2a}. The starting point was again $v_0=(0,0,0)$. However, instead of the null polarity defined above, the duality used to get the form diagram was
\begin{align}
(F_x,F_y,M,1)_{\sim} \mapsto (-F_y+1,F_x,-1,M)_{\sim}
\end{align}
which is no longer a polarity. It also maps $v_0=(0,0,0)$ to plane $z=x$ (plane $p_{0\sim}$). This means the moment diagram appears between planes $p_{0\sim}$ and $p_{i\sim}$, and the intersections of $p_{0\sim}$ and $p_{i\sim}$ would have to be projected down as well to get the lines of action of the resultants. The required projection is again an orthogonal vertical one, as it can be seen in Figure \ref{fig:pl2b}.

\begin{figure}[h]
\begin{subfigure}[t]{0.5\textwidth}
\centering
\includegraphics[scale=0.5]{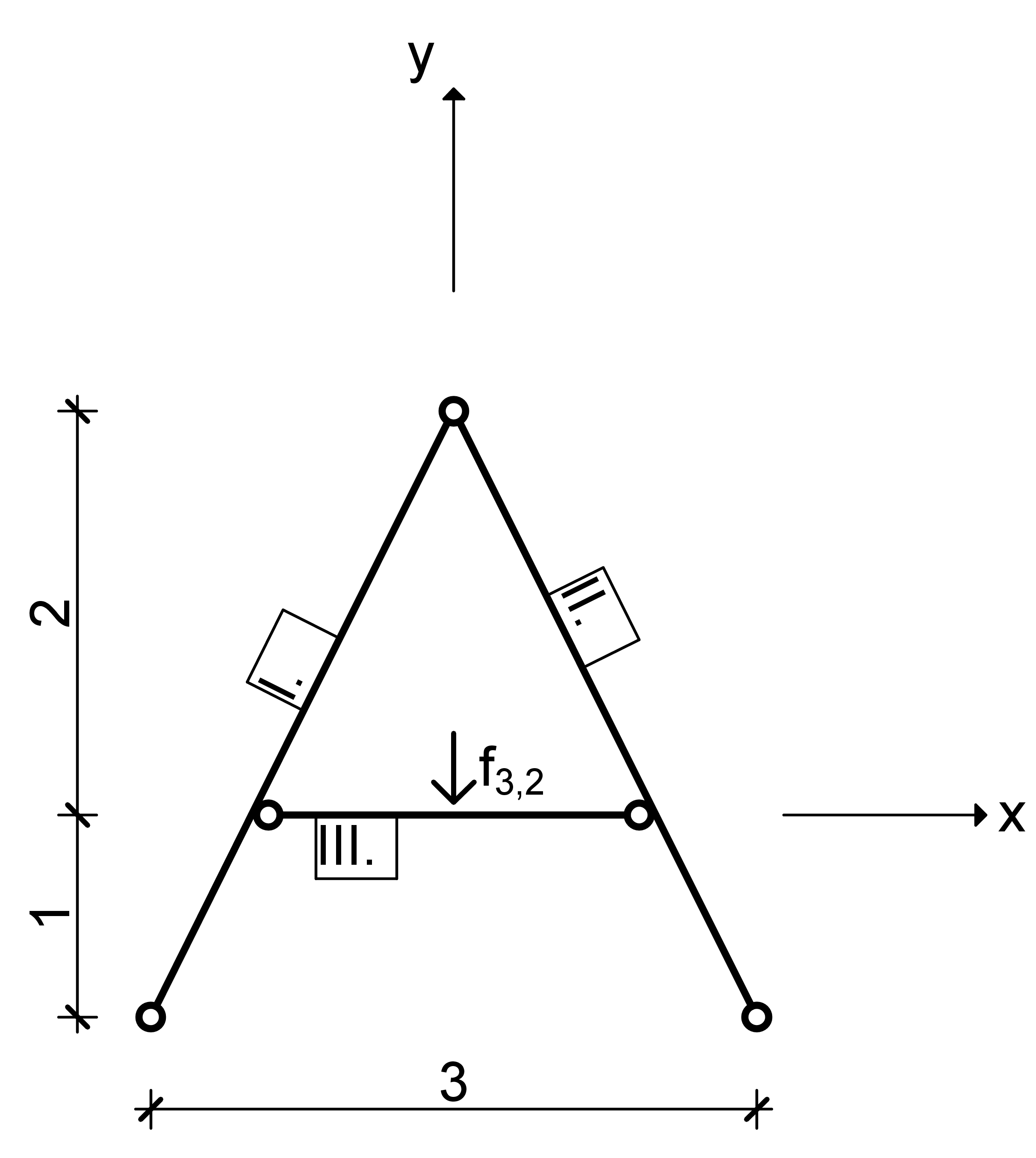}\\
\caption{Geometry of the third example}\label{fig:p3a}
\end{subfigure}
\begin{subfigure}[t]{0.5\textwidth}
\centering
\includegraphics[scale=0.5]{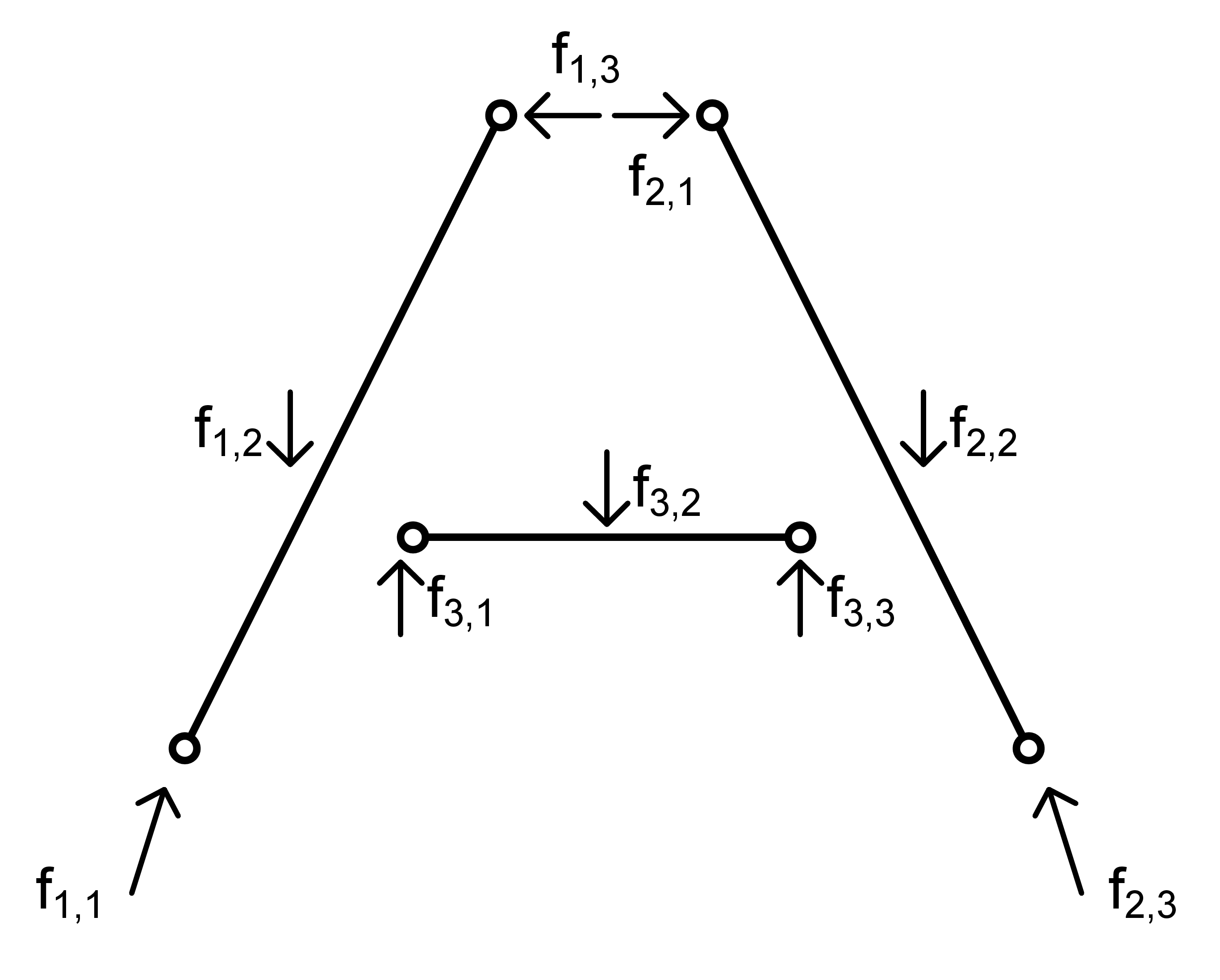}
\caption{Free-body diagram of the third example}\label{fig:p3b}
\end{subfigure}
\caption{}
\end{figure}

\paragraph{In the case of multiple bodies} we might not want to have the same starting point for each of the loops in the force diagram. To see such an example, consider the structure presented in Figure \ref{fig:p3a}, with the corresponding free-body diagram in Figure \ref{fig:p3b}. The indexing convention adopted was that $f_{i,j}$ is the $j-th$ force acting on body $i$. The statical equilibrium is satisfied with forces

\begin{align}
f_{1,1}=&(1/2,3,-5)\\
f_{1,2}=&(0,-3,3)=-f_{3,1}\\
f_{1,3}=&(-1/2,0,1)=-f_{2,1}\\
f_{2,2}=&(0,-3,-3)=-f_{3,3}\\
f_{2,3}=&(-1/2,3,5)\\
f_{3,2}=&(0,-6,0).
\end{align}

\begin{figure}[h]
\begin{subfigure}[h]{0.5\textwidth}
\includegraphics[width=1\textwidth]{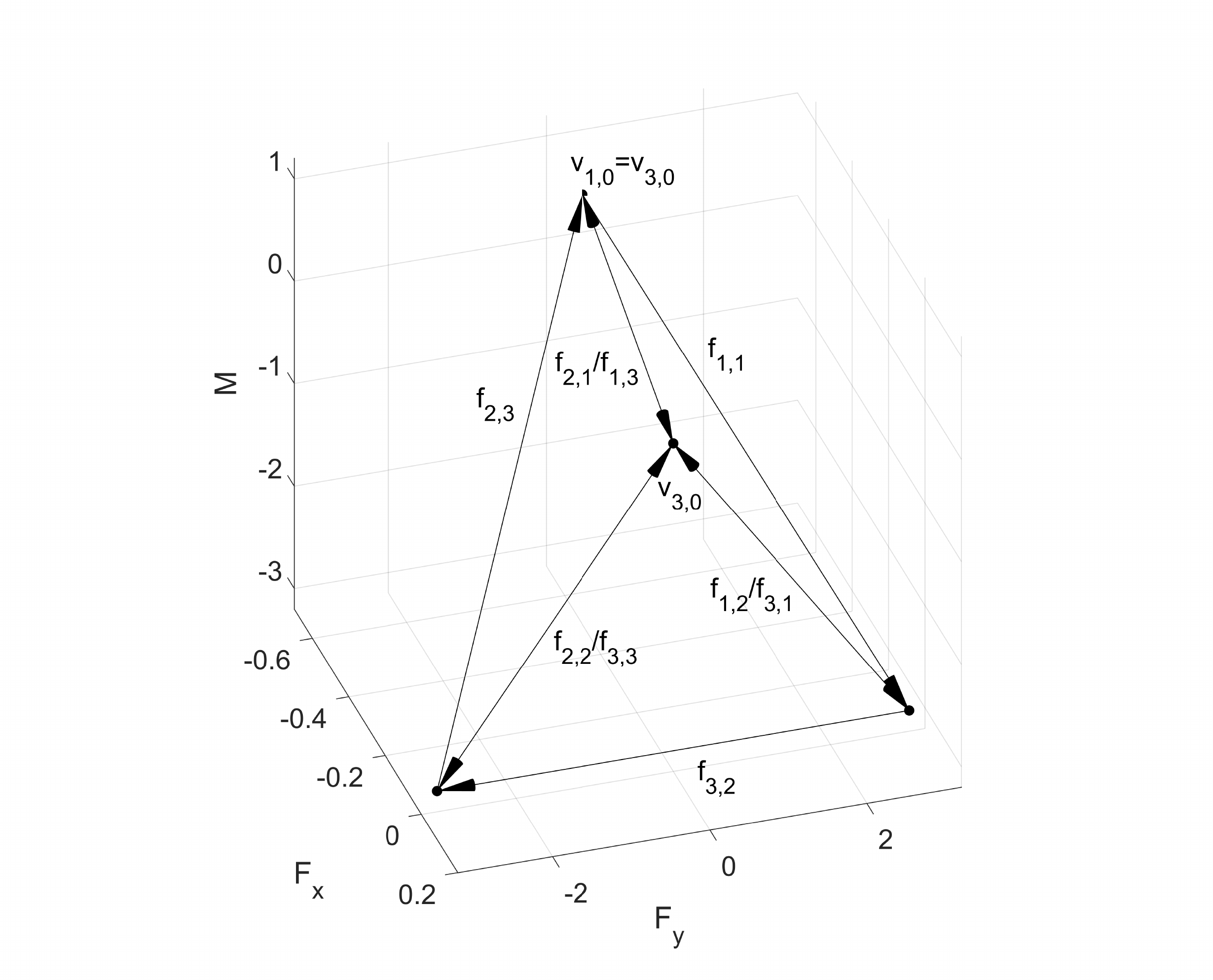}
\caption{Force diagram of the structure shown in Figure \ref{fig:p3a}. Note the different loops corresponding to the different bodies. Each loop is planar, since each set of forces meet at a single point.}\label{fig:pl3a}
\end{subfigure}
\begin{subfigure}[h]{0.5\textwidth}
\includegraphics[width=1\textwidth]{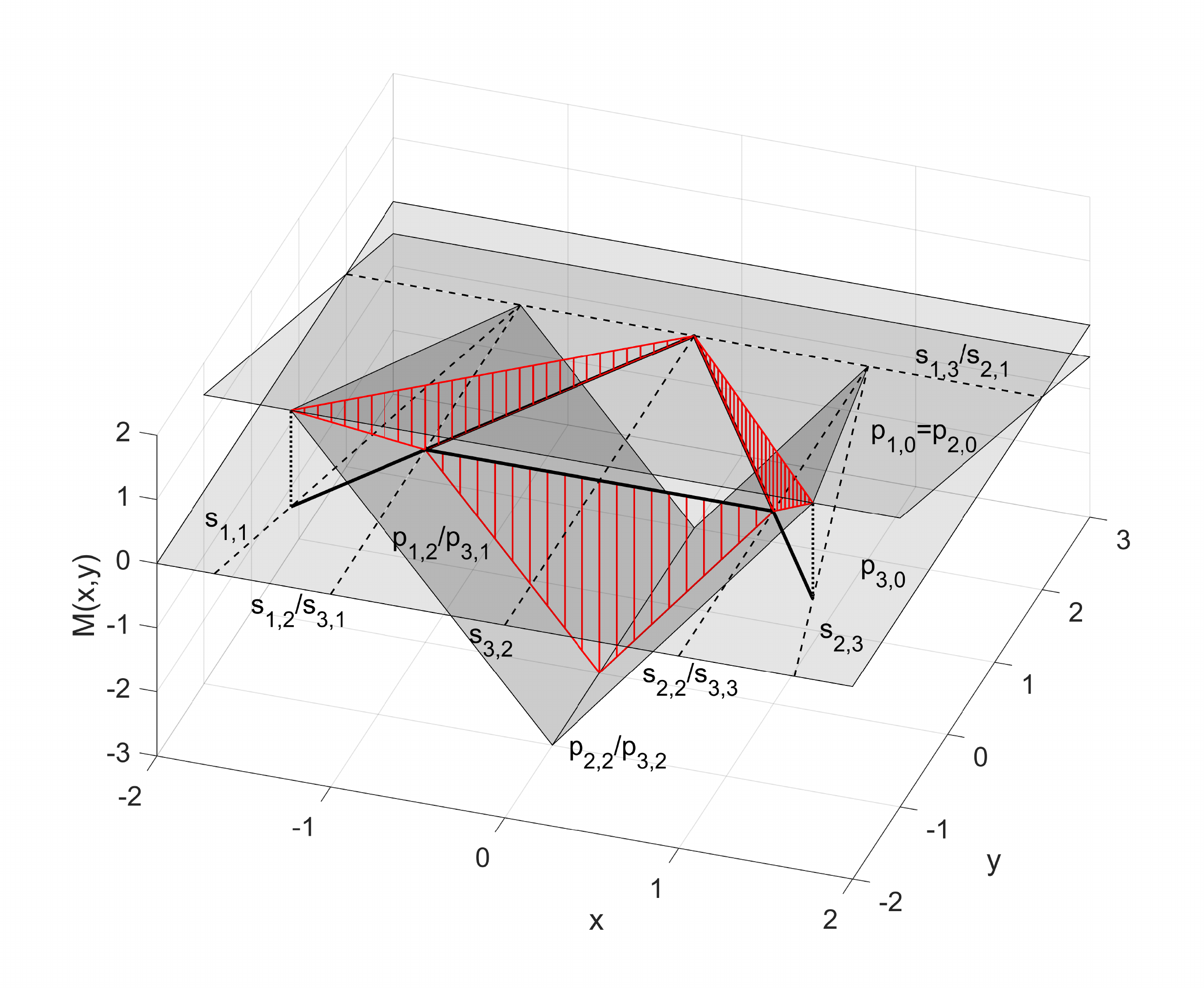}
\caption{Moment functionals evaluated above the structure given in Figure \ref{fig:p3a}. The moment diagram is shown in red. the lines of actions of the forces are denoted with $s_{i,j}$. Note the different reference planes $p_{1,0}=p_{2,0}$ and $p_{3,0}$.}\label{fig:pl3b}
\end{subfigure}
\caption{}
\end{figure}

Since all bodies are subjected to precisely 3 forces, the statical equilibrium requires these forces to meet at a single point, implying the planarity of each corresponding loop in the force diagram. We can arrange these loops similarly to Cremona's force plan, but in 3 dimensions, resulting in a polyhedron as presented in Figure \ref{fig:pl3a}. Each face of this polyhedron represents the equilibrium of a (sub-) body. We still have 3 degrees of freedom in deciding where to place the diagram (in this case $v_{3,0}=0$ was chosen), but the relative starting positions of the loops are determined by this construction. (In this case, we have $v_{1,0}=v_{2,0}=v_{3,1}-f_{1,1}$.) The duality used to get $p_i$ from $v_i$  was the same as in \eqref{eq:duC}, resulting in 
\begin{align}
p_{3,0}=(0,0,-1,0)_{\sim} \text{ \ and \ } p_{1,0}=p_{2,0}=(0,-1/2,-1,1)_{\sim}.
\end{align}
This means that while the moment diagram of the third body can be measured above or below the $z=0$ plane, the moment diagrams of the first and second bodies appear between the plane $y/2+z=1$ and  planes $p_{i,j}$ ($i,j\in \{1,2 \}$).

\section{The moment as a cross ratio}\label{sec:cr}
An earlier work \cite{2019arXivBaranyai} generalized Maxwell's duality-based method to include all projective dualities. The proposed method consisted of two polyhedra such that one is a dual of the other under some projective duality. The projected image (with some projection) of the edges of one of the polyhedra gave the 2D form diagram of the structure under consideration (which was a truss, but we see now how it need not necessarily be a truss). Similarly, a possibly different projection of the edges of the dual polyhedron provided the 2D force diagram (Cremona force-plan) of the structure. If we want to generalize this to our 3D force and 3D form diagrams, we can decompose arbitrary duality as the duality given in \eqref{eq:duC} and a pair of projective transformations $T_g$ and $T_f$, acting on the 3D form and 3D force diagrams respectively. The correctness of this approach was proven in the aforementioned paper. There the problem is reduced to multiple possible dualities instead of the one given in \eqref{eq:duC}, but with the help of affine transformations. Here the strict choice of \eqref{eq:duC} requires that $T_g$ and $T_f$ are more general, projective transformations.\\

Consider point $q_{\sim}$ of the structure, and let $m_{\infty\sim}$ denote the point at infinity in the vertical direction, which was the projection centre in the examples provided. Let us denote the intersection of line $\overleftrightarrow{q_{\sim},m_{\infty\sim}}$ with planes $p_{0\sim}$ and $p_{i\sim}$ with $m_{0\sim}$ and $m_{i\sim}$ respectively. As the examples have shown, the slice of the moment diagram corresponding to $q_{\sim}$ is the line segment $\overline{m_{0\sim}m_{i\sim}}$. By treating the moment functionals in a metric way, we implicitly imposed a scale on this line segment, which can be represented by adding point $m_{1\sim}$ to this line, at one unit distance from $m_{0\sim}$ in the positive direction. Since $\overline{m_{0\sim}m_{1\sim}}=1$ holds, we can express the value of the moment diagram as a cross ratio

\begin{align}
\overline{m_{0\sim}m_{i\sim}}=\frac{\overline{m_{0\sim}m_{i\sim}}}{\overline{m_{0\sim}m_{1\sim}}}=(m_{\infty\sim},m_{0\sim},m_{1\sim},m_{i\sim})
\end{align}
using that $m_{\infty\sim}$ is an ideal point. The importance of this is that the cross ratio is a projective invariant. Using the notation that $T_g(m_{\sim})=m'_{\sim}$, the value of the moment diagram can be calculated as the cross ratio $(m'_{\infty\sim},m'_{0\sim},m'_{1\sim},m'_{i\sim})$, which is the same as $\overline{m_{0\sim}m_{i\sim}}$ since projective transformations preserve cross ratios.\\ 
As for the 3D force diagrams, we will also measure moments as cross ratios on the projecting rays. Starting again with the examples provided, let $l_{\sim}$ be an arbitrary projecting ray (vertical line in this case). The fact that a force system $\sum f_i$ has moment $\sum M_i$ with respect to the origin can be expressed with the cross ratio 
\begin{align}
(l_{\infty\sim},l_{0\sim},l_{1\sim},l_{i\sim})=\sum M_i
\end{align}
where $l_{\infty\sim}$ is the ideal point of the projecting ray and $l_{0\sim}$, $l_{1\sim}$ and $l_{i\sim}$ are the horizontal projections of $v_0$, $v_0+(0,0,1)$ and $v_i$  to the projecting ray $l_{\sim}$. Since we are defining this value with a horizontal projection, it is easy to see how the choice of $l_{\sim}$ does not affect the value of the cross ratio. If we wish to use a general duality, we will have to construct, or we will get as a result a 3D force diagram that is a projective transform of the example diagram, under transformation $T_f$. The figure can be interpreted/created with the help of the cross ratio
\begin{align}
(T_f(l_{\infty\sim}),T_f(l_{0\sim}),T_f(l_{1\sim}),T_f(l_{i\sim}))=\sum M_i.
\end{align}

While at first this representation might look strange, in terms of actually representing a value in a planar image it is better then the area-based representational idea of the so-called Rankine-reciprocals. Since the cross ratio is projective-invariant, it is preserved during the operation in which we project the 3D figure to our 2D image plane (paper or computer screen). As such, given a ruler and a calculator it is possible to accurately reconstruct the values of the moment diagram from a single image. Contrast this with the theoretically impossible problem of determining and \emph{accurately} comparing the areas of polygons which were originally skew to the image plane.

\section{Relation to past works, summary}
In light of all this, we can understand Maxwell's \cite{maxwell1870} and Cremona's \cite{cremona1890graphical} graphical truss solving methods as a special case of this phenomenon. Special in the sense that the bars of the structure coincide with the lines of action of the forces, and special in the sense that the force diagrams of the forces acting on a joint form planar polygons. (Since all forces $f_i$ acting on a joint have to pass through the location of the joint, say $(j_x,j_y)$, we have $\scal{s_i}{(j_x,j_y,1)}=0$ for all $i$ implying this.) These planar polygons formed the faces of the polyhedra Maxwell related to what is now called Airy stress function. This brings us to the works of Hegedüs \cite{hegedus85} and Williams and McRobie \cite{williams16}, who found moment diagrams while creating discontinuous stress functions to analyse discontinuous structures. We have shown where these diagrams come from and how these stress-functions can be considered as the evaluation of the appropriate moment functionals. Also, the choice of $v_0$ or $p_0$, carries the same $3$ degrees of freedom as the stress function has: due to differentiating twice, stress functions $F(x,y)$ and $F(x,y)+ax+by+c \quad  (a,b,c\in\mathbb{R})$ produce the same stresses. With this, we have seen the underlying connection between different past approaches and fully understood the reason why they work.\\

Furthermore, it was also shown how to interpret the moment values in the generalized version \cite{2019arXivBaranyai} of Maxwell's reciprocal construction. The approach provided here can also be useful for modern graphical analysis of structures, with the newly proposed method to construct the line of action of the resultant from the (3 dimensional) force diagram. \\

We have also seen how the need for a 3 dimensional duality to pass from vector to functional form arises from the 3 dimensionality of the forces involved. Correspondingly, the generalization of this method to 3 dimensional objects loaded with arbitrary forces remains an interesting open question, as one might attempt a 6 dimensional approach. There is a proposition to use 4 dimensional polyhedra\cite{mcrobie2017geometry} present in current literature, which is tied to Rankine's representational idea. While the higher dimensional approach can be simpler (the Dandelin spheres, usually also presented in projective geometry classes are a classic example of this), the representation of these 6 components raises a few, interesting questions.

\section*{Appendix}
\label{sec:appa}
Consider a vector-space $\mathbb{V}$ over the real numbers, and the following equivalence relation
\begin{align}
x\sim y \iff y\in\{\lambda x  \ \vert \ \lambda \in \mathbb{R}\setminus \{0\} \} \quad (x,y\in \mathbb{V}).\label{eq:equivalence}
\end{align}
The set of all such $y$ is called the equivalence class of $x$ (denoted with $x_{\sim}$), while the vectors themselves are called representants of the equivalence class. If $\mathbb{V}$ is $n$ dimensional, we can consider each equivalence class as a point in an $n-1$ dimensional projective space. How one embeds the affine space into the projective one  carries certain freedom in terms of coordinates. In this paper point $\vec{x}\in \mathbb{R}^{n-1}$ is mapped to the equivalence class $(\vec{x},1)\sim $. In keeping with the duality principle of projective geometry, hyperplanes of the projective space can also be thought of as equivalence classes, such that hyperplane $h_{\sim}$ contains point $x_\sim$, if and only if $\scal{h}{x}=0$ holds. 

\bibliographystyle{unsrt} 
\bibliography{agsbib}

\end{document}